\documentclass[runningheads]{llncs}
\usepackage[T1]{fontenc}
\usepackage{cite}
\usepackage{float}
\usepackage{hyperref}
\usepackage{graphicx}

\title{Mi-Go: Test Framework which uses YouTube as Data Source for Evaluating Speech Recognition Models like OpenAI's Whisper}

\author{Tomasz Wojnar\inst{1}\orcidID{0000-0003-0389-7645} \and
Jarosław Hryszko\inst{1}\orcidID{0000-0002-4207-1080} \and
Adam Roman\inst{1}\orcidID{0000-0002-1020-5128}}

\authorrunning{T. Wojnar et al.}
\titlerunning{Mi-Go: A Test Framework for Speech Recognition Models}

\institute{Jagiellonian University, Faculty of Mathematics and Computer Science
\\Division of Software Engineering
\\Łojasiewicza 6, 30-348 Krakow, Poland
\\ \email{tomasz.wojnar@student.uj.edu.pl}, \email{\{jaroslaw.hryszko, adam.roman\}@ii.uj.edu.pl}\\
}

\begin{document}
\maketitle

\begin{abstract}
This article introduces Mi-Go, a novel testing framework aimed at evaluating the performance and adaptability of general-purpose speech recognition machine learning models across diverse real-world scenarios. The framework leverages YouTube as a rich and continuously updated data source, accounting for multiple languages, accents, dialects, speaking styles, and audio quality levels. To demonstrate the effectiveness of the framework, the Whisper model, developed by OpenAI, was employed as a test object. The tests involve using a total of 124 YouTube videos to test all Whisper model versions. The results underscore the utility of YouTube as a valuable testing platform for speech recognition models, ensuring their robustness, accuracy, and adaptability to diverse languages and acoustic conditions. Aditionally, by contrasting the machine-generated transcriptions against human-made subtitles, the Mi-Go framework can help pinpoint potential misuse of YouTube subtitles, like Search Engine Optimization.
\end{abstract}

\section{Introduction}

Speech recognition has become a critical component in numerous applications, ranging from virtual assistants and transcription services to voice-controlled devices and accessibility tools. The increasing reliance on speech recognition machine learning models necessitates robust and comprehensive evaluation methodologies to ensure their performance, reliability, and adaptability across diverse scenarios. 

Existing test frameworks for speech recognition models often rely on curated datasets, such as LibriSpeech~\cite{panayotov2015librispeech}, CommonVoice~\cite{ardila2019common}, and TIMIT~\cite{zue1990speech}. While these datasets provide a controlled environment for evaluation, they may not capture the full spectrum of real-world scenarios, potentially limiting the model's generalizability. Additionally, these datasets may not be updated frequently, resulting in potential stagnation in performance evaluation.

In this article, we introduce Mi-Go (the name will be explained further), 
a test framework designed to assess the performance of general-purpose speech recognition machine learning models. Mi-Go harnesses the power of YouTube as a data source, providing access to a virtually unlimited repository of diverse audio-visual content. YouTube offers a rich and continuously updated collection of spoken language data, encompassing various languages, accents, dialects, speaking styles, and audio quality levels. This makes it an ideal platform for testing the adaptability and performance of speech recognition models in real-world situations.

In recent years, there has been a growing interest in harnessing the vast amount of data available on platforms such as YouTube for machine learning tasks. Various approaches have been proposed to collect and process data from YouTube, including YouTube-8M~\cite{abu2016youtube} and AudioSet~\cite{gemmeke2017audio}. However, these methods primarily focus on video and audio classification tasks rather than the evaluation of speech recognition models.

The landscape of speech recognition technology has witnessed a paradigm shift, driven by rapid advancements in deep learning and artificial intelligence. Groundbreaking architectures, such as recurrent neural networks (RNNs), convolutional neural networks (CNNs), and, more recently, transformer-based models, have revolutionized this domain, offering unprecedented accuracy in transcribing human speech. These models, trained on vast datasets, have demonstrated remarkable proficiency in navigating the complexities of language, including accents, dialects, and noise interference. The emergence of these models not only underscores the accelerated pace of development in this field, but also leads one to believe that in the near future seamless human-computer interaction will become the norm.
It should be noted that while these advancements present exciting prospects, they also raise compelling questions concerning data privacy, algorithmic bias, and the digital divide.

In our study, we address this need by proposing, and then empirically investigating the effectiveness 
of speech recognition model test framework, which utilize YouTube as a data source, providing access to an extensive and diverse collection of audio samples for evaluation purposes. This approach ensures that the performance assessment remains up-to-date and relevant, capturing the nuances of real-world speech more accurately than curated data sets. To the best of our knowledge, there is no research on using YouTube videos -- and video subtitles provided by the YouTube users -- for speech recognition evaluation. Considering all the above,
our goal is to answer the following research question:

\begin{itemize}
    \item[(RQ)] \textbf{Will testing the selected speech recognition machine learning model using YouTube as a data source, as made possible by Mi-Go, produce similar results (measured using the same metric) as the tests conducted by the model creators?}
\end{itemize}

Mi-Go automates the process of data extraction, annotation, and evaluation from YouTube, ensuring an up-to-date and representative sample for testing purposes. By leveraging algorithms for data filtering and annotation, Mi-Go facilitates a thorough and unbiased evaluation of the speech recognition models. Moreover, Mi-Go is designed to be easily adaptable, allowing for seamless integration with variety of different speech recognition solutions, making it a versatile and valuable tool in the speech recognition research community.

The primary motivation behind the development of the Mi-Go test framework stems from the recognition of several limitations in existing approaches to evaluate speech recognition models. As speech recognition technology continues to play a critical role in various applications, including voice assistants, transcription services, and accessibility tools, ensuring the robustness and accuracy of these models is crucial.

Traditional test frameworks often rely on static, curated datasets which, while useful for establishing a controlled environment, may not fully represent the diversity and complexity of real-world speech scenarios. This can lead to overfitting and limit the model's generalizability, ultimately affecting its performance in real-world applications.

Additionally, as the field of speech recognition rapidly advances, existing evaluation methods
may struggle to keep pace with new developments and challenges, potentially hindering the progress of these models. By utilizing YouTube as a data source, Mi-Go aims to overcome these limitations and offers a more comprehensive and dynamic evaluation environment.

Another motivation for the development of Mi-Go is the need for a flexible and adaptable framework capable of accommodating variety of speech recognition models. This adaptability allows researchers and developers to compare and contrast the performance of various models, facilitating the continuous improvement and refinement of speech recognition systems.

By addressing these limitations and providing a dynamic, diverse, and adaptable testing framework, Mi-Go aspires to contribute significantly to the field of speech recognition research, driving innovation and fostering the development of highly accurate and robust models for various applications.

\section{YouTube as a Platform for Speech Recognition Model Testing} 
With over 2 billion monthly active users and a diverse array of content uploaded every day, YouTube offers a rich resource for researchers and developers working on speech recognition technology. By tapping into this wealth of multilingual and multi-genre content, it is possible to evaluate and refine speech recognition models across various languages, dialects, and acoustic environments.

\textbf{Diversity of Content}. YouTube's vast library of user-generated content covers an extensive range of topics, languages, and styles. This diversity enables the evaluation of speech recognition models in real-world scenarios, such as noisy environments, various accents, and even low-quality audio recordings. By testing models on such a diverse dataset, researchers can identify potential weaknesses and areas for improvement, ultimately resulting in more robust and accurate speech recognition systems.

\textbf{Multilingual Corpus}. One of the key advantages of using YouTube for speech recognition model testing is the platform's multilingual nature. Videos on the site are available in numerous languages, allowing for the assessment of models' performance across different linguistic settings. This multilingual corpus is invaluable for developing models that can handle a variety of languages, accents, and dialects, thereby expanding their utility and applicability.

\textbf{Availability of Human-Generated Transcripts}. Many YouTube videos come with human-generated subtitles, either provided by content creators or contributed by users through the platform's community contributions feature. These transcripts serve as valuable ground-truth data for evaluating speech recognition models, as they offer a reliable source of comparison for the models' output. By comparing model-generated transcriptions with human-generated ones, researchers can assess the accuracy and performance of their models, identifying areas where improvements are needed.

\textbf{Potential for Continuous Model Improvement}. The ever-growing volume of content on YouTube presents an opportunity for continuous improvement and adaptation of speech recognition models. As new videos are uploaded, models can be retested and fine-tuned to ensure they remain up-to-date and effective in an ever-changing linguistic landscape. This continuous feedback loop helps researchers identify trends, challenges, and emerging language patterns, which can be incorporated into model updates.

YouTube is an invaluable platform for speech recognition model testing due to its diverse, multilingual content and the availability of human-generated transcripts. By leveraging this vast resource, researchers and developers can evaluate and refine their models, ensuring they are robust, accurate, and adaptable to a variety of languages and acoustic conditions.

\section{Speech recognition model used}

OpenAI, a company most notably recognized for its contribution to the field of artificial intelligence through the development of advanced large language models like GPT-3 and GPT-4, also developed state-of-the-art, general-purpose speech recognition model, which demonstrate exceptional performance in various applications, called Whisper~\cite{radford2022robust}. 
Due to proven outstanding performance of that model, as well as fact, that it has been made available under a open-source (MIT) licence, we decided, that first application of Mi-Go framework will be evaluation of the Whisper model, using YouTube-derived data. Through our research, we aim to contribute to the ongoing efforts in improving the performance, reliability, and adaptability of speech recognition models in real-world scenarios.

At this point we should explain, that the name ,,Mi-Go'' comes from a novella by H.P. Lovecraft called ,,The Whisperer in Darkness''; thus, in our opinion, it would make a good name for the tool primarily used to test the Whisper model.

The model is based on a Transformer sequence-to-sequence architecture and is trained on a range of speech processing tasks, including multilingual speech recognition, speech translation, spoken language identification, and voice activity detection. These tasks are collectively represented as a sequence of tokens to be predicted by the decoder, enabling a single model to supplant multiple stages of a conventional speech-processing pipeline. The multitask training approach employs a series of unique tokens that act as task specifiers or classification targets~\cite{radford2022robust}.


Whisper model is available in five different sizes. Four of them (tiny, base, small, medium) having additional English-only versions, which -- according to the creators -- perform better when used in English-only applications~\cite{WhisperReadme}. Thus, in our research, we decided to use English-only model versions.
Each model offers a balance between speed and accuracy. The names of the used models, their approximate memory requirements and relative speeds are provided in Table~\ref{tab:whisper-models}.

\begin{center}
\begin{table}[!ht]
    \caption{Comparison of Whisper models~\cite{WhisperReadme}}
    \centering
    \begin{tabular}{|l|c|c|c|c|}
    \hline
    Name of the model & Number of parameters & Layers & Relative speed & Required VRAM \\ \hline
    tiny.en              & 39000000  & 4 & $\sim$ 32x & $\sim$ 1 GB    \\ 
    base.en              & 74000000 & 6 &  $\sim$ 16x & $\sim$ 1 GB \\ 
    small.en             & 244000000 & 12 & $\sim$ 6x & $\sim$ 2 GB    \\ 
    medium.en            & 769000000  & 24 & $\sim$ 2x & $\sim$ 5 GB    \\ 
    large             & 1550000000 & 32 & $\sim$ 1x & $\sim$ 10 GB \\ \hline
    \end{tabular}
    \label{tab:whisper-models}
\end{table}
\end{center}

\section{Mi-Go framework}

Framework proposed in this paper, which allows for testing speech recognition machine learning models (like Whisper) was written in Python programing language. Its source code is available for download under Apache-2.0 license at the following address:

\url{https://github.com/Kowalski1024/Mi-Go}

Mi-Go framework consists of, apart from various small supporting modules, the following three major components (we will describe them further):
\begin{itemize}
    \item \emph{TestPlan Generator},
    \item \emph{(YouTube) TestRunner},
    \item \emph{TranscriptTest (TranscriptDifference)}.
\end{itemize}

The \emph{TestPlan Generator} is a component that creates test plan in the form of JSON file, tailored for a specific test data source. Users decide what the test plan should include, via command line interface.
Current version of \emph{TestPlan Generator} is designed to use YouTube videos as test data. It utilizes the YouTube Data API to search for the videos by according to the given filters and the external Python library called \emph{youtube-transcript-api}\footnote{Available from: https://pypi.org/project/youtube-transcript-api/, access: 28.07.2023} to retrieve information about available video transcripts. Users can choose filter such like language, category, duration or how many videos it should contain. 

The JSON file contains all the necessary metadata about the videos used in further testing, it also preserves information about the selected filters and token for YouTube Data API to create another collection of test data (videos, subtitles), if needed.


\begin{figure}[!ht]
\centering
\includegraphics[width=\textwidth]{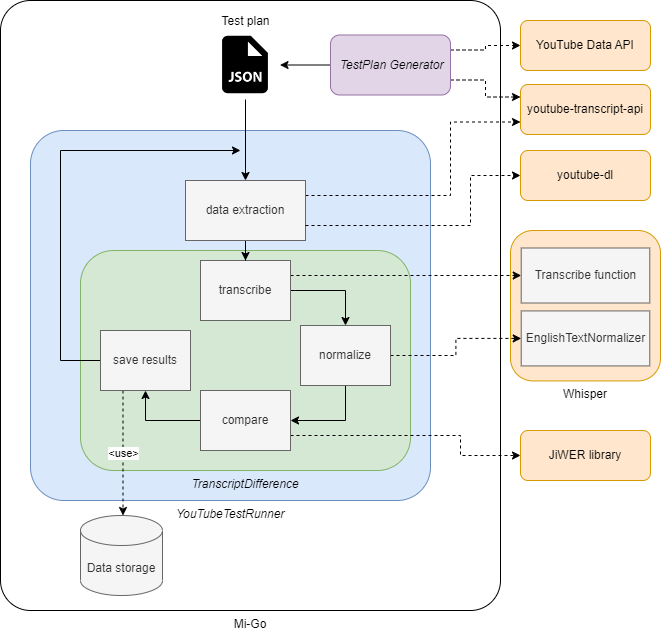}
\caption{\label{fig:general_idea}General idea of Mi-Go tool.}
\end{figure}


The \emph{TestRunner} is a component designed to execute tests, basing on a specific test plan provided by \emph{TestPlan Generator} in a form of JSON file. The TestRunner download and processes the test data, performs the tests by calling \emph{TranscriptTest} component (described below) and return the results. A ,,decorator'' design pattern \emph{register} is featured within the \emph{TestRunner}, enabling users to effortlessly add new transcript tests to the class, making them easily selectable via command-line arguments.

For the sake of flexibility, \emph{TestRunner} component can be easily extended by additional, external code allowing it to work with a specific test data source -- in case of using YouTube, we extended this component and called it \emph{YouTubeTestRunner}. It uses the test plan produced by the \emph{TestPlan Generator} and prepares the test data by downloading audio using \emph{youtube-dl} and transcript text (subtitles) from YouTube by \emph{youtube-transcript-api}. Then it performs tests and results are saved in an SQLite database and also in a JSON file. This file has the same structure as the test plan (but with results included), can be used as a test plan to generate further test iterations -- for example, to create a new set of videos with given parameters, as we described earlier.

The \emph{TranscriptTest} 
is a component that executes the speech recognition machine learning model against audio data collected from test data source and compares model's output (text) with the original, man-made transcript text also collected from test data source.

\emph{TranscriptTest} component can be adjusted for specified speech recognition model by extending that component with model-specific code. As we consider Whisper model, Whisper-specific code is provided by \emph{TranscriptDifference} component. That component, before calling \emph{TranscriptTest}, normalizes test data (transcript texts) using a built-in Whisper normalization function\footnote{Refer to: https://github.com/openai/whisper/blob/main/whisper/normalizers/english.py, access: 28.07.2023} and uses a open-source JiWER library\footnote{Available from: https://github.com/jitsi/jiwer, access: 30.05.2023 } to calculate WER speech recognition metrics (described in Section~\ref{sec:metrics}).

\section{Experimental setup}
Here we describe an experimental setup that leverages the Mi-Go Framework to use YouTube videos, across all categories, as a test data to test and evaluate speech recognition models by comparing their output with human-made transcripts. The purpose of the experiment is to confirm whether the following setup (Mi-Go and YouTube as test data source) will allow to test the speech recognition Whisper model and obtain test results similar to those obtained by the Whisper model creators.

\begin{figure}[!ht]
\centering
\includegraphics[width=\textwidth]{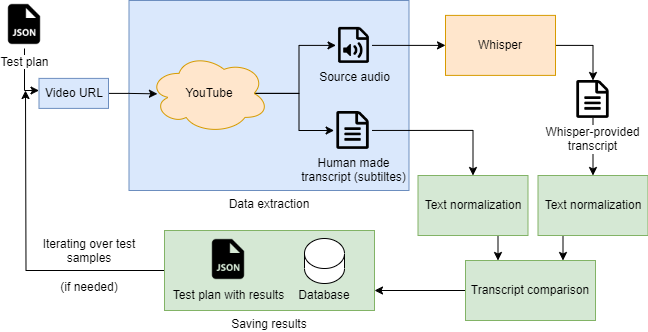}
\caption{\label{fig:experiment_pipeline}Experiment pipeline.}
\end{figure}

\subsection{Data Collection and Preparation}\label{sec:data_collection}
To begin the experimental setup, we instruct the \emph{TestPlan Generator}, a component of the Mi-Go Framework, via command line interface, to  to fetch \textbf{8-10 videos} per category listed in Table~\ref{tab:category}.

\begin{table}[!ht]
    \caption{YouTube videos categories considered in the experiment}   
    \centering
    \begin{tabular}{|l|c|}
    \hline
         Category & Number of videos fetched \\ \hline
         Autos \& Vehicles & 10 \\ 
         Comedy & 10 \\ 
         Education & 9 \\ 
         Entertainment & 9 \\ 
         Film \& Animation & 9 \\ 
         Howto \& Style & 10 \\ 
         Music & 9 \\ 
         News \& Politics & 10 \\ 
         Nonprofits \& Activism & 10 \\ 
         People \& Blogs & 8 \\ 
         Pets \& Animals & 10\\ 
         Science \& Technology & 10 \\ 
         Travel \& Events & 10 \\ \hline
    \end{tabular}
 \label{tab:category}
\end{table}

These videos are selected basing on factors such as \textbf{popularity, relevance, and the presence of human-generated subtitles}, ensuring a diverse and high-quality dataset. The YouTube Data API is used to acquire the videos, while the \emph{youtube-transcript-api} library retrieves their corresponding transcripts. Already fetched, the same set of videos is used to test all versions of Whisper model (listed in Table~\ref{tab:whisper-models}). Full list of 124 videos used in experiment is provided in Appendix~\ref{app:videos}.

\subsection{Test Plan Generation}
Once the set of videos is compiled, the \emph{TestPlan Generator} creates a JSON file containing the test plans tailored for the \emph{YouTube Runner}, another component of the Mi-Go Framework. The test plans include information about the videos, such as their URLs, categories, and other metadata.

\subsection{Test Execution and metrics}\label{sec:metrics}
The \emph{YouTube Runner} is responsible for executing the tests on the collected videos. It prepares the test data by extracting audio and subtitles, as specified in the test plans generated by the \emph{TestPlan Generator}. The \emph{TranscriptDifference} 
component is then utilized to run the speech recognition model (OpenAI's Whisper in this case) on the audio files, comparing model's output (generated text) with the human-generated transcripts. This process allows for an evaluation of the models' effectiveness across all YouTube categories.

For that evaluation, similarly to Whisper model creators~\cite{radford2022robust}, Mi-Go framework uses Word Error Rate (WER) measure. WER is a common metric used to assess the performance of speech recognition systems, automatic translation systems, and other tasks involving transcription or translation. It is calculated by determining the minimum number of operations needed to transform the system output into the correct output. These operations include (see Equation~\ref{eq:wer}): word insertions $I$, word deletions $D$, and word substitutions $S$. To compute the WER, the total number of these operations is divided by the total number of words in the correct output $N$ (in our case: total number of words in subtitles attached to a particular YouTube video), yielding a ratio that represents the rate of errors per word. The lower the WER, the better the performance of the system, as it means fewer errors were made.

\begin{equation}
\label{eq:wer}
\mbox{WER} = \frac{S + D + I}{N}
\end{equation}

The concept of WER has been part of the field of automatic speech recognition and computational linguistics for many years. It is based on the Levenshtein distance or edit distance, a string metric for measuring the difference between two sequences, introduced by Vladimir Levenshtein in 1965~\cite{levenshtein1965binary}. The exact individual or group that first applied this concept specifically as Word Error Rate in speech recognition or translation systems is not clearly documented. It likely emerged from the academic and industry communities working on speech and language processing technologies. WER has since become a standard measure in these fields. In some cases, WER is expressed as a percentage (by multiplying the  original formula by 100\%), especially when easy understanding of the measure is a main concern. In our reserach we kept WER in its original form for the sake of simplicity.

\subsection{Results Storage and Analysis}
The test results are stored in an SQLite database. Additionally, the results are saved in a JSON file with the same structure as the original test plans, allowing for easy generation of the next test plan iteration. This dual storage approach facilitates simple access, filtering, and analysis of the test results.

\section{Results}

To answer research question, we used proposed Mi-Go framework, to utilize~124 YouTube videos, representing all categories listed in Table~\ref{tab:category}, to test all five available Whisper models (as listed in Table~\ref{tab:whisper-models}) and collect Word Error Rate metrics as a result.


Statistics for collected Word Error Rate values for all tested models are presented in Table~\ref{tab:all-results}. Detailed statistics of the WER value for each model by category, are presented in the appendix~A. 



\begin{table}[!ht]
    \caption{Word Error Rate value statistics for all tested model versions}
    \centering
    \begin{tabular}{|l|c|c|c|c|c|c|}
    \hline
    Model version & Min    & Mean   & Median  & Max  & Std. deviation  & Variance \\ \hline
    tiny.en       & 0.014  & 0.506  & 0.122   & 18.81 & 1.74           & 3.028 \\
    base.en       & 0.009  & 2.377  & 0.102   & 253  & 22.699          & 515.242 \\
    small.en      & 0.011  & 0.495  & 0.08    & 20.474 & 1.932         & 3.734 \\
    medium.en     & 0.006  & 0.583  & 0.079   & 22.4 & 2.178           & 4.745 \\
    \textbf{large}         & \textbf{0.005}  & \textbf{0.276}  & \textbf{0.081}   & \textbf{3.987} & \textbf{0.521}          & \textbf{0.272} \\
    Aggregate results  & 0.005  & 0.848  & 0.093   & 253  & 10.271          & 105.489 \\ \hline
        
    \end{tabular}
    \label{tab:all-results}
\end{table}

Whisper model characteristics, published by its authors, concern only 'large' model -- thus, in Table~\ref{tab:all-results} we presented WER statistics for that model with bold font.

\begin{figure}[ht!]
\centering
\includegraphics[width=\textwidth]{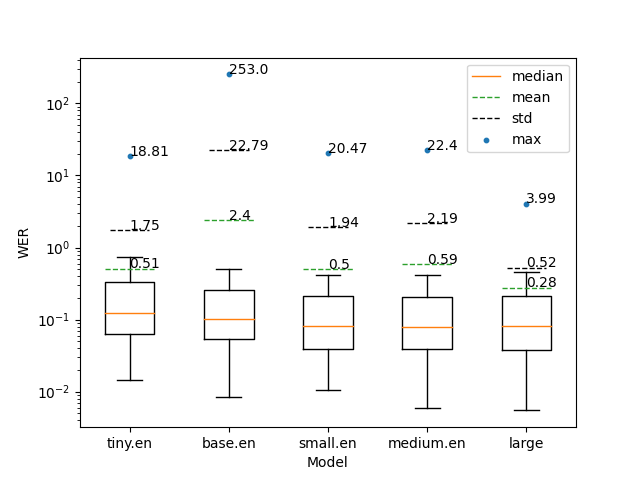}
\caption{Box plot of experiment results. Note the logarithmic scale.}\label{fig:boxplot}
\end{figure}

As we can see, the median for 'large' model test results is WER = 0.081. The worst median of results for Whisper 'large' model presented by its creators was 19.6~\cite{radford2022robust}. That result was achieved by using CORAAL speech recording dataset popularized by Gunter et al.~\cite{gunter2021contextualizing}. Other datasets used to validate Whisper model by its creators were: recordings of earning calls by Del Rio et al.~\cite{del2021earnings}, sets of recordings of online blogs and podcasts, and dataset containing recordings of The Late Show (sic!). These Whisper 'large' model evaluation results from~\cite{radford2022robust} compared to our results, are presented in Table~\ref{tab:comparison}. By making this comparison, we can conclude, that YouTube-based Whisper model tests done by us produce similar results as the tests conducted by the Whisper model creators, using different test data.

\begin{table}[!ht]
    \caption{Comparison of WER median values presented in~\cite{radford2022robust} and our results (highlighted), using YouTube as a dataset source.}
    \centering
    \begin{tabular}{|l|c|}
    \hline
        Dataset used & WER median \\ \hline
        TED-LIUM3 & 0.035 \\ 
        Meanwhile & 0.051 \\ 
        \textbf{YouTube} & \textbf{0.081} \\ 
        Kincaid46 & 0.088 \\ 
        Earnings-21 & 0.097 \\ 
        Rev16 & 0.113 \\ 
        Earnings-22 & 0.126 \\ 
        CORAAL & 0.196 \\ \hline
    \end{tabular}
    \label{tab:comparison}
\end{table}

One can find large WER values for few results, significantly different from the median (as illustrated in Figure~\ref{fig:boxplot}). However, by reviewing the YouTube videos that were used for the tests that ended with high WER values, we can conclude that the reason for this is not due to a malfunction of the Mi-Go tool or Whisper model. Instead, the high WER values are due to the actual discrepancy between the human-made subtitles attached to the video and those generated by the Whisper model. We found that such discrepancies occur due to several reasons:

\begin{enumerate}
    \item \textbf{Transcription errors}. Humans, despite their proficiency, are not infallible and may make mistakes when transcribing speech to text. This could involve mishearing words or phrases, particularly in a noisy environment, during rapid speech, or when dealing with dialectal variations or accents.
    
    \item \textbf{Interpretation differences}. Subtitling is not always a direct one-to-one transcription process. The transcriber's understanding and interpretation of the speech can influence the outcome. Homonyms, idiomatic expressions, cultural references, or ambiguous statements can all be interpreted differently depending on the transcriber's knowledge and perspective.
    
    \item \textbf{Contextual adaptations}. Subtitle makers often make deliberate changes to the text for various reasons. They may simplify or clarify speech to make it more accessible to the audience, especially if the speech is complex or jargon-filled. They may also modify the text to match reading speed constraints, given that text must be readable within the time it is displayed. Cultural adaptations may also be made to make the content more comprehensible to a specific audience.
    
    \item \textbf{Descriptive transcriptions}. Some transcriptions go beyond the spoken content and provide descriptions of the visual elements in the video. These are often intended for visually impaired or blind viewers to provide them a more comprehensive understanding of the video content. Such case occured with video resulted with highest WER value in our experiment (WER = 253). While that video only consist of animal sounds, actual subtitles are as follows (original spelling)\footnote{https://www.youtube.com/watch?v=4Co4mDeCIJ4, access: 30.05.2023}:
    
    \begin{quote}
        Cats Cats are very cute animals Animals that are close and affectionate with people Cat breed is a species with relatively high fertility, giving birth to 2-3 litters of kittens a year New born kittens only weighs about 100g and fits easily in the palm of your hand Horses are smart, wise animals Mother horses as young as 3 years old can start breeding (...)
    \end{quote}
    
    \item \textbf{Search engine optimization (SEO)}. Some subtitles may be created or modified with the goal of improving the video's visibility in search engine results. The inclusion of relevant keywords and phrases can make the video more likely to appear in search results related to those terms, hence enhancing the video's discoverability. Here is an example of such subtitles from one of videos we used\footnote{https://www.youtube.com/watch?v=JbioSJtMkwQ, access: 30.05.2023}:
    
    \begin{quote}
        The Animals,Funniest Animals Video,Funny Video,Funny Animals,Cats, Dogs,Funny Cats,Funny Dogs,Pets,Funny Pets,Funny,Cute,Cute Animals,Cute Pets,Funny Cat Video,Funny Dog Video,Funny Animals Life,Wow,Best Animals,Best Animals Video,Compilation,Funny Video Compilation,Kittens,Puppies,Try not to laugh,Best Animals 2023,Best of 2022,Cute Puppy,Funny Kitten,Animals International,Funny Animal Video.
    \end{quote}
    
\end{enumerate}

By comparing model-made transcription to the existing human-made subtitles, discrepancies can be identified. Factors such as background noise, speaker accents, or low-quality audio can impact the model's performance. Hence, although speech recognition models can help identify potential inaccuracies in subtitles, a degree of human oversight and validation is typically necessary to confirm and rectify these inaccuracies. From a different perspective, automated setup which utilize Mi-Go and Whisper, \textbf{can significantly help in detection of video subtitles misuse.}

\section{Conclusions and Future work}


In this paper, we have introduced Mi-Go, a novel, lightweight, and flexible test framework for evaluating general-purpose speech recognition models, specifically targeting the Whisper model and using YouTube's vast and diverse content. Traditional evaluation methods, which employ curated datasets, may not capture the broad array of real-world scenarios, hence potentially limiting a model's generalizability. Mi-Go, by leveraging YouTube's dynamic content, offers an enriched platform for testing these models. It incorporates a Test Plan Generator and a TestRunner script, streamlining the testing process while providing detailed results in an SQLite database and a JSON file for easy access and analysis. The OpenAI-developed model, Whisper, was used as a test object. 
With a transformer-based architecture and training on various speech processing tasks, Whisper represents a versatile tool for speech recognition tasks. Testing was performed on 124 YouTube videos, demonstrating the usefulness of the Mi-Go framework in assessing model performance and identifying discrepancies between model-generated transcriptions and human-made subtitles. The results underscore the necessity for human oversight in rectifying inaccuracies and the potential of the Mi-Go framework for enhancing speech recognition models' robustness and adaptability. 

While the Mi-Go test framework demonstrates promising results in evaluating speech recognition models like Whisper, several avenues for future work can further enhance its capabilities:

\begin{enumerate}
\item Expanding the framework to accommodate other data sources (like non-English YouTube videos or video hosting services other than YouTube), providing an even more diverse and representative set of audio samples for evaluation.
\item Incorporating advanced techniques for data preprocessing and augmentation, which can help in simulating various real-world challenges, such as background noise and audio distortions.
\item Developing a user-friendly interface or API, making it easier for researchers and developers to integrate and utilize the Mi-Go framework in their projects.
\item Extending the framework to support other tasks, such as speaker identification and language identification, in addition to speech recognition.
\end{enumerate}

Nevertheless, we hope that in its current form, Mi-Go tool will find wide application in both speech recognition machine learning model testing and detection of anomalies in existing video transcriptions.

\newpage

\bibliographystyle{splncs04}
\bibliography{bibliography}

\newpage

\appendix

\section{Detailed results for particular models, broken down by YouTube categories}\label{app:detailed_results}

\begin{table}[H]
    \centering
    \begin{tabular}{|l|l|l|l|l|l|l|}
    \hline
        \textbf{Min} & \textbf{Max} & \textbf{Mean} & \textbf{Std. deviation} & \textbf{Variance} & \textbf{Median} & \textbf{Category} \\ \hline

        0.025 & 1.016 & 0.189 & 0.317 & 0.1 & 0.056 & Autos \& Vehicles \\ 
        0.028 & 0.152 & 0.102 & 0.039 & 0.002 & 0.101 & Comedy \\ 
        0.015 & 0.115 & 0.05 & 0.031 & 0.001 & 0.036 & Education \\ 
        0.069 & 0.966 & 0.5 & 0.367 & 0.135 & 0.573 & Entertainment \\ 
        0.038 & 1 & 0.582 & 0.396 & 0.157 & 0.565 & Film \& Animation \\ 
        0.013 & 0.115 & 0.053 & 0.034 & 0.001 & 0.045 & Howto \& Style \\ 
        0.083 & 3.987 & 1.099 & 1.41 & 1.987 & 0.458 & Music \\ 
        0.051 & 0.677 & 0.254 & 0.18 & 0.032 & 0.186 & News \& Politics \\ 
        0.005 & 0.125 & 0.038 & 0.035 & 0.001 & 0.031 & Nonprofits \& Activism \\ 
        0.025 & 0.228 & 0.099 & 0.07 & 0.005 & 0.097 & People \& Blogs \\ 
        0.053 & 1.134 & 0.63 & 0.481 & 0.232 & 0.952 & Pets \& Animals \\ 
        0.021 & 0.18 & 0.057 & 0.046 & 0.002 & 0.044 & Science \& Technology \\ 
        0.015 & 0.123 & 0.045 & 0.031 & 0.001 & 0.037 & Travel \& Events \\ \hline
    \end{tabular}
    \caption{Word Error Rate statistics for model 'large'}
\end{table}

\begin{table}[H]
    \centering
    \begin{tabular}{|l|l|l|l|l|l|l|}
    \hline
        \textbf{Min} & \textbf{Max} & \textbf{Mean} & \textbf{Std. deviation} & \textbf{Variance} & \textbf{Median} & \textbf{Category} \\ \hline
        0.028 & 4.787 & 0.636 & 1.499 & 2.247 & 0.059 & Autos \& Vehicles \\ 
        0.03 & 0.254 & 0.099 & 0.065 & 0.004 & 0.079 & Comedy \\ 
        0.009 & 0.121 & 0.041 & 0.034 & 0.001 & 0.032 & Education \\ 
        0.057 & 5.68 & 1.467 & 1.916 & 3.67 & 0.959 & Entertainment \\ 
        0.045 & 3.792 & 1.202 & 1.412 & 1.994 & 0.958 & Film \& Animation \\ 
        0.016 & 0.132 & 0.066 & 0.043 & 0.002 & 0.069 & Howto \& Style \\ 
        0.069 & 22.4 & 2.937 & 7.32 & 53.576 & 0.29 & Music \\ 
        0.053 & 0.681 & 0.249 & 0.18 & 0.032 & 0.191 & News \& Politics \\ 
        0.006 & 0.106 & 0.04 & 0.033 & 0.001 & 0.033 & Nonprofits \& Activism \\ 
        0.03 & 0.157 & 0.093 & 0.052 & 0.003 & 0.095 & People \& Blogs \\ 
        0.031 & 3.26 & 0.879 & 0.973 & 0.946 & 0.985 & Pets \& Animals \\ 
        0.022 & 0.173 & 0.056 & 0.044 & 0.002 & 0.045 & Science \& Technology \\ 
        0.02 & 0.12 & 0.048 & 0.029 & 0.001 & 0.041 & Travel \& Events \\ \hline
    \end{tabular}
    \caption{Word Error Rate statistics for model 'medium'}
\end{table}

\begin{table}[H]
    \centering
    \begin{tabular}{|l|l|l|l|l|l|l|}
    \hline
        \textbf{Min} & \textbf{Max} & \textbf{Mean} & \textbf{Std. deviation} & \textbf{Variance} & \textbf{Median} & \textbf{Category} \\ \hline
        0.02 & 4.456 & 0.58 & 1.39 & 1.931 & 0.054 & Autos \& Vehicles \\ 
        0.037 & 0.222 & 0.101 & 0.054 & 0.003 & 0.089 & Comedy \\ 
        0.013 & 0.121 & 0.047 & 0.033 & 0.001 & 0.035 & Education \\ 
        0.048 & 20.474 & 3.028 & 6.6 & 43.559 & 1.028 & Entertainment \\ 
        0.044 & 3.259 & 1.083 & 1.02 & 1.041 & 0.994 & Film \& Animation \\ 
        0.015 & 0.113 & 0.056 & 0.037 & 0.001 & 0.05 & Howto \& Style \\ 
        0.078 & 1.202 & 0.34 & 0.357 & 0.128 & 0.162 & Music \\ 
        0.05 & 0.683 & 0.252 & 0.179 & 0.032 & 0.198 & News \& Politics \\ 
        0.011 & 0.115 & 0.039 & 0.033 & 0.001 & 0.029 & Nonprofits \& Activism \\ 
        0.026 & 0.288 & 0.125 & 0.091 & 0.008 & 0.12 & People \& Blogs \\ 
        0.052 & 3.213 & 0.864 & 0.953 & 0.908 & 0.987 & Pets \& Animals \\ 
        0.02 & 0.171 & 0.055 & 0.043 & 0.002 & 0.042 & Science \& Technology \\ 
        0.015 & 0.125 & 0.049 & 0.03 & 0.001 & 0.041 & Travel \& Events \\ \hline
    \end{tabular}
    \caption{Word Error Rate statistics for model 'small'}
\end{table}

\begin{table}[H]
    \centering
    \begin{tabular}{|l|l|l|l|l|l|l|}
    \hline
        \textbf{Min} & \textbf{Max} & \textbf{Mean} & \textbf{Std. deviation} & \textbf{Variance} & \textbf{Median} & \textbf{Category} \\ \hline
        0.03 & 5.263 & 0.756 & 1.666 & 2.775 & 0.073 & Autos \& Vehicles \\ 
        0.059 & 0.253 & 0.136 & 0.072 & 0.005 & 0.114 & Comedy \\ 
        0.023 & 0.125 & 0.056 & 0.03 & 0.001 & 0.053 & Education \\ 
        0.057 & 2.108 & 1.014 & 0.874 & 0.764 & 0.995 & Entertainment \\ 
        0.046 & 2.347 & 0.87 & 0.716 & 0.513 & 0.95 & Film \& Animation \\ 
        0.018 & 0.117 & 0.061 & 0.04 & 0.002 & 0.047 & Howto \& Style \\ 
        0.17 & 0.694 & 0.379 & 0.211 & 0.044 & 0.254 & Music \\ 
        0.06 & 0.687 & 0.266 & 0.177 & 0.031 & 0.224 & News \& Politics \\ 
        0.009 & 0.121 & 0.043 & 0.035 & 0.001 & 0.031 & Nonprofits \& Activism \\ 
        0.037 & 0.273 & 0.15 & 0.097 & 0.009 & 0.164 & People \& Blogs \\ 
        0.067 & 253 & 25.872 & 79.806 & 6369.016 & 0.984 & Pets \& Animals \\ 
        0.028 & 0.183 & 0.066 & 0.045 & 0.002 & 0.057 & Science \& Technology \\ 
        0.021 & 0.139 & 0.065 & 0.032 & 0.001 & 0.057 & Travel \& Events \\ \hline
    \end{tabular}
    \caption{Word Error Rate statistics for model 'base'}
\end{table}

\begin{table}[H]
    \centering
    \begin{tabular}{|l|l|l|l|l|l|l|}
    \hline
        \textbf{Min} & \textbf{Max} & \textbf{Mean} & \textbf{Std. deviation} & \textbf{Variance} & \textbf{Median} & \textbf{Category} \\ \hline
        0.037 & 2.382 & 0.383 & 0.736 & 0.542 & 0.087 & Autos \& Vehicles \\ 
        0.096 & 0.336 & 0.177 & 0.075 & 0.006 & 0.147 & Comedy \\ 
        0.033 & 0.132 & 0.063 & 0.03 & 0.001 & 0.058 & Education \\ 
        0.064 & 18.81 & 2.991 & 6.004 & 36.05 & 0.998 & Entertainment \\ 
        0.07 & 1.717 & 0.95 & 0.624 & 0.389 & 1 & Film \& Animation \\ 
        0.023 & 0.125 & 0.072 & 0.041 & 0.002 & 0.065 & Howto \& Style \\ 
        0.267 & 1.309 & 0.64 & 0.396 & 0.157 & 0.541 & Music \\ 
        0.076 & 0.693 & 0.275 & 0.177 & 0.031 & 0.226 & News \& Politics \\ 
        0.014 & 0.12 & 0.054 & 0.038 & 0.001 & 0.042 & Nonprofits \& Activism \\ 
        0.064 & 0.384 & 0.222 & 0.118 & 0.014 & 0.241 & People \& Blogs \\ 
        0.089 & 1.87 & 0.801 & 0.685 & 0.469 & 0.985 & Pets \& Animals \\ 
        0.036 & 0.201 & 0.077 & 0.05 & 0.002 & 0.06 & Science \& Technology \\ 
        0.03 & 0.146 & 0.083 & 0.036 & 0.001 & 0.072 & Travel \& Events \\ \hline
    \end{tabular}
    \caption{Word Error Rate statistics for model 'tiny'}
\end{table}

\newpage

\section{List of YouTube videos used for speech recognition model testing}\label{app:videos}

\subsection{Autos \& Vehicles}

\begin{enumerate}
\item \url{https://www.youtube.com/watch?v=sM0znZ_Ur8E}
\item \url{https://www.youtube.com/watch?v=rDg-yena4Ec}
\item \url{https://www.youtube.com/watch?v=HpxdGq0NECw}
\item \url{https://www.youtube.com/watch?v=0UplB_26F3Y}
\item \url{https://www.youtube.com/watch?v=EidxYt03zeM}
\item \url{https://www.youtube.com/watch?v=1Ja9gzsh8So}
\item \url{https://www.youtube.com/watch?v=583OJ0F31lM}
\item \url{https://www.youtube.com/watch?v=RkbzA-0VnU4}
\item \url{https://www.youtube.com/watch?v=VvjpX_p5cXU}
\item \url{https://www.youtube.com/watch?v=bwoHmubxDIg}
\end{enumerate}

\subsection{Comedy}

\begin{enumerate}
\item \url{https://www.youtube.com/watch?v=xCyebt737pU}
\item \url{https://www.youtube.com/watch?v=i7nmnWXQgVw}
\item \url{https://www.youtube.com/watch?v=7aEqxAxwRuw}
\item \url{https://www.youtube.com/watch?v=asY7CQJkUa8}
\item \url{https://www.youtube.com/watch?v=Vx1-HS3q0wQ}
\item \url{https://www.youtube.com/watch?v=hdPCNwXE3GU}
\item \url{https://www.youtube.com/watch?v=Bq7O57JOFAM}
\item \url{https://www.youtube.com/watch?v=V-9pGpCgoGQ}
\item \url{https://www.youtube.com/watch?v=4TIIdrOfbls}
\item \url{https://www.youtube.com/watch?v=b277LPZyJUk}
\end{enumerate}

\subsection{Education}

\begin{enumerate}
\item \url{https://www.youtube.com/watch?v=-TrHHGKQxcI}
\item \url{https://www.youtube.com/watch?v=wX78iKhInsc}
\item \url{https://www.youtube.com/watch?v=y3fm6wNzK70}
\item \url{https://www.youtube.com/watch?v=S294zRodS_4}
\item \url{https://www.youtube.com/watch?v=7n2hCebmT4c}
\item \url{https://www.youtube.com/watch?v=cGkQil14LPQ}
\item \url{https://www.youtube.com/watch?v=rhgwIhB58PA}
\item \url{https://www.youtube.com/watch?v=g1pb2aK2we4}
\item \url{https://www.youtube.com/watch?v=w5zV_GusqvQ}
\end{enumerate}

\subsection{Entertainment}

\begin{enumerate}
\item \url{https://www.youtube.com/watch?v=A9yhyB9VSsc}
\item \url{https://www.youtube.com/watch?v=4pJBUAymAU8}
\item \url{https://www.youtube.com/watch?v=aF57wvoVplg}
\item \url{https://www.youtube.com/watch?v=JkSB4eOBe6w}
\item \url{https://www.youtube.com/watch?v=Mc-WiSHXq1g}
\item \url{https://www.youtube.com/watch?v=K_WZo6Yfl9Q}
\item \url{https://www.youtube.com/watch?v=fYfFBrROcxM}
\item \url{https://www.youtube.com/watch?v=_PfWOgVtijc}
\item \url{https://www.youtube.com/watch?v=z2WclC-UubY}
\end{enumerate}

\subsection{Film \& Animation}

\begin{enumerate}
\item \url{https://www.youtube.com/watch?v=kNw8V_Fkw28}
\item \url{https://www.youtube.com/watch?v=AZS5cgybKcI}
\item \url{https://www.youtube.com/watch?v=3Aq6UtLKtzM}
\item \url{https://www.youtube.com/watch?v=1D7D3_HFB3o}
\item \url{https://www.youtube.com/watch?v=D_Rx4qZ8QRc}
\item \url{https://www.youtube.com/watch?v=8uj0Vy8oMkw}
\item \url{https://www.youtube.com/watch?v=a2wpvc96i24}
\item \url{https://www.youtube.com/watch?v=t023ryQgguQ}
\item \url{https://www.youtube.com/watch?v=gZyjJtBIlow}
\end{enumerate}

\subsection{Howto \& Style}

\begin{enumerate}
\item \url{https://www.youtube.com/watch?v=Pp_7rK7BJ5Q}
\item \url{https://www.youtube.com/watch?v=6QKZM3EZ4ng}
\item \url{https://www.youtube.com/watch?v=VQeNtw3SHqo}
\item \url{https://www.youtube.com/watch?v=WEGmOnpOvRM}
\item \url{https://www.youtube.com/watch?v=kUE2fPLOUxo}
\item \url{https://www.youtube.com/watch?v=-eqcnPq2xdE}
\item \url{https://www.youtube.com/watch?v=XmQ8mZFqczw}
\item \url{https://www.youtube.com/watch?v=n781HdPa7pA}
\item \url{https://www.youtube.com/watch?v=6xl0fGqsUA4}
\item \url{https://www.youtube.com/watch?v=BY3sQ-mZgqs}
\end{enumerate}

\subsection{Music}

\begin{enumerate}
\item \url{https://www.youtube.com/watch?v=b1kbLwvqugk}
\item \url{https://www.youtube.com/watch?v=WcIcVapfqXw}
\item \url{https://www.youtube.com/watch?v=Y2NkuFIlLEo}
\item \url{https://www.youtube.com/watch?v=hT_nvWreIhg}
\item \url{https://www.youtube.com/watch?v=fV4DiAyExN0}
\item \url{https://www.youtube.com/watch?v=f74GYIVMk3I}
\item \url{https://www.youtube.com/watch?v=Soa3gO7tL-c}
\item \url{https://www.youtube.com/watch?v=D2KE2a5qo0g}
\item \url{https://www.youtube.com/watch?v=Uq9gPaIzbe8}
\end{enumerate}

\subsection{News \& Politics}

\begin{enumerate}
\item \url{https://www.youtube.com/watch?v=bb1WJquq8jI}
\item \url{https://www.youtube.com/watch?v=2k6_7NJsX4k}
\item \url{https://www.youtube.com/watch?v=4d0ife1L3mM}
\item \url{https://www.youtube.com/watch?v=uiwxkvgno1k}
\item \url{https://www.youtube.com/watch?v=LQzsQU7_hVw}
\item \url{https://www.youtube.com/watch?v=cjVoABxV0NA}
\item \url{https://www.youtube.com/watch?v=iF0MLrl44sc}
\item \url{https://www.youtube.com/watch?v=GPirQlhXCls}
\item \url{https://www.youtube.com/watch?v=cdjScH9BLVw}
\item \url{https://www.youtube.com/watch?v=OxrfdrEO1m4}
\end{enumerate}

\subsection{Nonprofits \& Activism}

\begin{enumerate}
\item \url{https://www.youtube.com/watch?v=UzdF2zpex8o}
\item \url{https://www.youtube.com/watch?v=UaDqc1dT7Rc}
\item \url{https://www.youtube.com/watch?v=mrPjz30rAVQ}
\item \url{https://www.youtube.com/watch?v=5IlKJGV_Z_8}
\item \url{https://www.youtube.com/watch?v=Z-6IfEoETyU}
\item \url{https://www.youtube.com/watch?v=KEoxUw-gwec}
\item \url{https://www.youtube.com/watch?v=JyPf7uoCGyE}
\item \url{https://www.youtube.com/watch?v=Yt38f7A_Rwo}
\item \url{https://www.youtube.com/watch?v=3m6OGbLTQgY}
\item \url{https://www.youtube.com/watch?v=cSvKk0iySnU}
\end{enumerate}

\subsection{People \& Blogs}

\begin{enumerate}
\item \url{https://www.youtube.com/watch?v=lj5GXZaE7qs}
\item \url{https://www.youtube.com/watch?v=MujDIvNQQoI}
\item \url{https://www.youtube.com/watch?v=waL3eqJBwIk}
\item \url{https://www.youtube.com/watch?v=7AeMhVN-TFA}
\item \url{https://www.youtube.com/watch?v=WgPZt7WGZJk}
\item \url{https://www.youtube.com/watch?v=Ij7KyhaBdJ8}
\item \url{https://www.youtube.com/watch?v=h1yvhYsueAk}
\item \url{https://www.youtube.com/watch?v=bcb0Ig_Jv5E}
\end{enumerate}

\subsection{Pets \& Animals}

\begin{enumerate}
\item \url{https://www.youtube.com/watch?v=JbioSJtMkwQ}
\item \url{https://www.youtube.com/watch?v=ABgQYcyhCgg}
\item \url{https://www.youtube.com/watch?v=Dl9Sa4H5TM0}
\item \url{https://www.youtube.com/watch?v=ZtasaWN6WaU}
\item \url{https://www.youtube.com/watch?v=mKoF48g89s4}
\item \url{https://www.youtube.com/watch?v=j0SF0A6aDOU}
\item \url{https://www.youtube.com/watch?v=geIWxM7QvKE}
\item \url{https://www.youtube.com/watch?v=lGjSbHgX4jk}
\item \url{https://www.youtube.com/watch?v=wRpvm3B5Ocg}
\item \url{https://www.youtube.com/watch?v=4Co4mDeCIJ4}
\end{enumerate}

\subsection{Science \& Technology}

\begin{enumerate}
\item \url{https://www.youtube.com/watch?v=Oa9aWdcCC4o}
\item \url{https://www.youtube.com/watch?v=5pVjCJDAyhk}
\item \url{https://www.youtube.com/watch?v=V1HHvnd22lQ}
\item \url{https://www.youtube.com/watch?v=OyQ3B1U8_XY}
\item \url{https://www.youtube.com/watch?v=t7RaVnEGkc0}
\item \url{https://www.youtube.com/watch?v=dhq2WPLKiuY}
\item \url{https://www.youtube.com/watch?v=TYuJdnn6NyE}
\item \url{https://www.youtube.com/watch?v=SEI0LtUmpn4}
\item \url{https://www.youtube.com/watch?v=6ulwahtVQAU}
\item \url{https://www.youtube.com/watch?v=ONs9FCY74p0}
\end{enumerate}

\subsection{Travel \& Events}

\begin{enumerate}
\item \url{https://www.youtube.com/watch?v=kFMHx6XwBk0}
\item \url{https://www.youtube.com/watch?v=VxMHHqSOSTk}
\item \url{https://www.youtube.com/watch?v=WLSnrXEtrT4}
\item \url{https://www.youtube.com/watch?v=RB1MN0QoXH0}
\item \url{https://www.youtube.com/watch?v=Wt4XODPm4hA}
\item \url{https://www.youtube.com/watch?v=Cqawiry5HTY}
\item \url{https://www.youtube.com/watch?v=0RoA5QEm8fc}
\item \url{https://www.youtube.com/watch?v=9wbNabuP6aM}
\item \url{https://www.youtube.com/watch?v=dBdthTOx9YU}
\item \url{https://www.youtube.com/watch?v=dNU1lJiDaSY}
\end{enumerate}

\end{document}